\documentclass{kluwer}    
\usepackage{graphicx}

\newdisplay{guess}{Conjecture}

\begin{document}                                                                                   
\begin{article}
\begin{opening}         
\title{XTE\,J1550$-$564: A superluminal ejection during the September 1998
outburst}
\author{Diana \surname{Hannikainen}}
\runningauthor{Hannikainen et al.}
\runningtitle{XTE\,J1550$-$564}
\institute{Department of Physics \& Astronomy, University of Southampton,
Southampton SO17~1BJ, UK}
\author{Duncan \surname{Campbell-Wilson}, Richard \surname{Hunstead},
    Vince \surname{McIntyre}\thanks{Present address: ATNF,
    CSIRO, PO Box 76, Epping, NSW 1710, Australia}}
\institute{School of Physics, University of Sydney, NSW 2006, Australia}
\author{Jim \surname{Lovell}, John \surname{Reynolds}, Tasso \surname{Tzioumis}}
\institute{ATNF, CSIRO, PO Box 76, Epping, NSW 1710, Australia}
\author{Kinwah \surname{Wu}\thanks{Permanent address:
    School of Physics, University of Sydney, NSW 2006, Australia}}
\institute{MSSL, University College London, Holmbury St. Mary, Dorking,
    Surrey, RH5~6NT, UK}

\begin{abstract}
In 1998 September, the X-ray transient XTE\,J1550$-$564 underwent a
major outburst in soft and hard X-rays, followed by a radio flare. 
Australian Long Baseline Array images
obtained shortly after the peak in the radio flare showed evolving structure.
The components observed have an apparent separation velocity of $>2c$.
\end{abstract}

\end{opening}           

\section{Introduction}
The X-ray transient XTE\,J1550$-$564 was discovered simultaneously by
   the All-Sky Monitor on the Rossi X-ray Timing Explorer \cite{smith} and by
   the Burst and Transient Source Experiment on the Compton Gamma-Ray
   Observatory \cite{wilson} on MJD~51063 (1998~Sept~7; MJD$=$JD$-$240000.5).
Following the discovery of the radio counterpart \cite{cw},
   XTE\,J1550$-$564 was monitored with the Molonglo Observatory Synthesis
   Telescope (MOST) at 843~MHz between MJD~51065 and 51092.
A flare was recorded, peaking with an estimated flux density of 380~mJy
   on MJD~51078.
Two Very Long Baseline Interferometry (VLBI) observations using the
Australian Long Baseline Array (LBA) were performed 2--3
   days after the MOST flare.
In this paper we present the resulting images from these observations, and
   discuss briefly the implications of the evolving structure observed.

\section{Observations and results}
The VLBI observations were carried out on 1998 Sept 24 and 25 and are
   summarized in Table~\ref{tabobs}. The data post-correlation used the
   {\sc aips} task {\sc fring} to find fringes and calibrate the fringe
   amplitude.
Due to the small number of baselines, the Caltech program {\sc difmap} was used
   to model and source fit the data, and measure the flux densities.

\begin{table}
\caption{VLBI Observations}
\label{tabobs}
\begin{tabular}{lllllll} \hline
Date         & & T$_{\rm start}$ & T$_{\rm end}$ & &Antenna     & Frequency \\
         &    &  (UT) & (UT) & & & (GHz) \\ \hline
1998 Sept 24 & & 02:17           & 06:08         & & DSS45 (34m) & 2.29 \& 8.4\\
\multicolumn{2}{l}{Mid-point (MJD):}& \multicolumn{2}{l}{51080.7} & &
Hobart (26m) & 2.29 \& 8.4\\             &  &                 &               &
& Mopra (22m)  & 2.29   \\  &              &                 &              & &
& \\

1998 Sept 25 &  & 22:30           & 09:00         & & DSS43 (70m) &  2.29 \& 8.4\\
\multicolumn{2}{l}{Mid-point (MJD):}& \multicolumn{2}{l}{51081.7}& &
Hobart (26m)&  2.29 \& 8.4\\           &   &                &               & &
Mopra (22m)         &  2.29 \\           &   &                &               &
& ATCA (6$\times$22m) & 2.29\\ \hline
\end{tabular}
\end{table}

\begin{figure} 
\centerline{\includegraphics[width=21pc,angle=270]{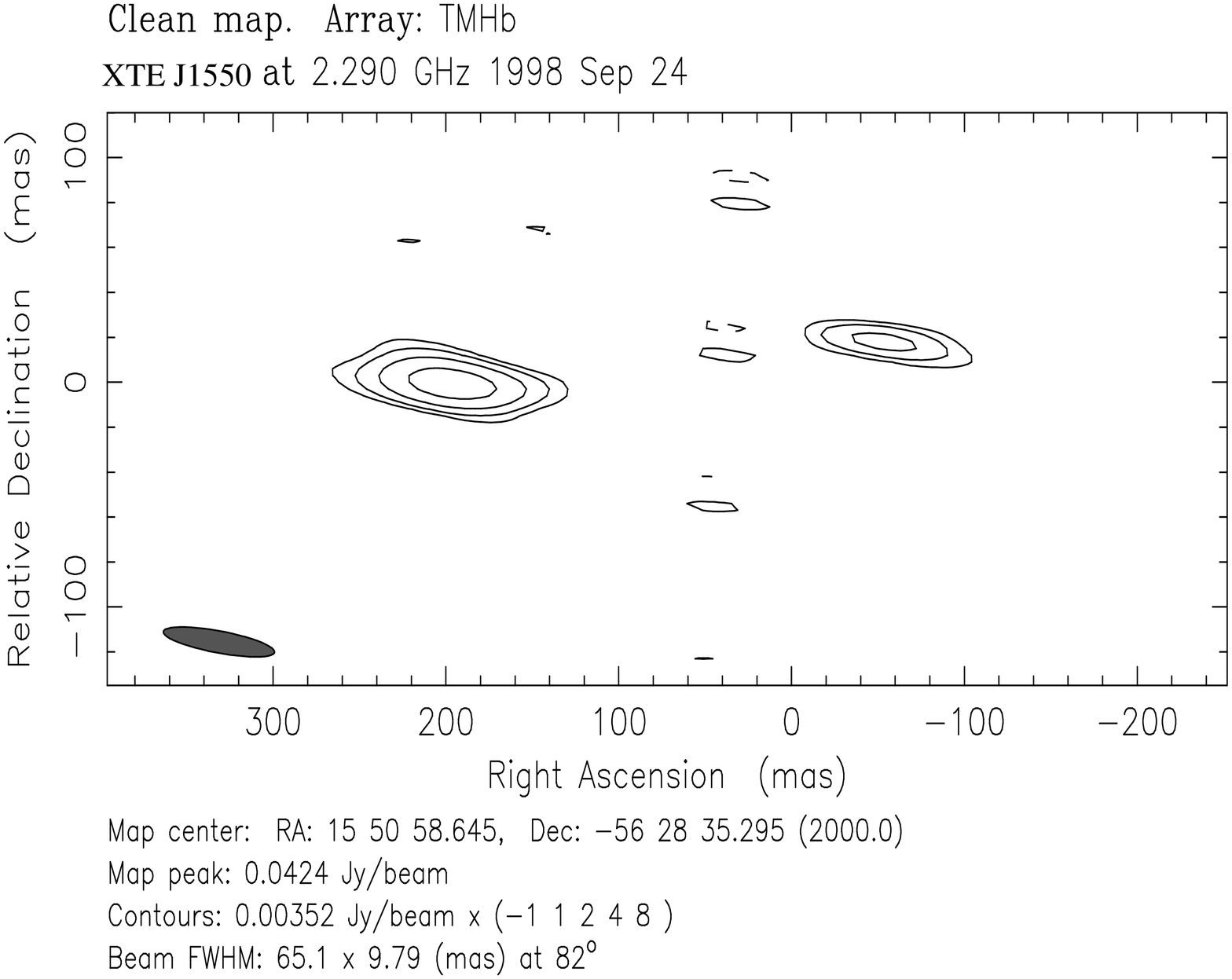}}
\vspace{0.75cm}
\centerline{\includegraphics[width=21pc,angle=270]{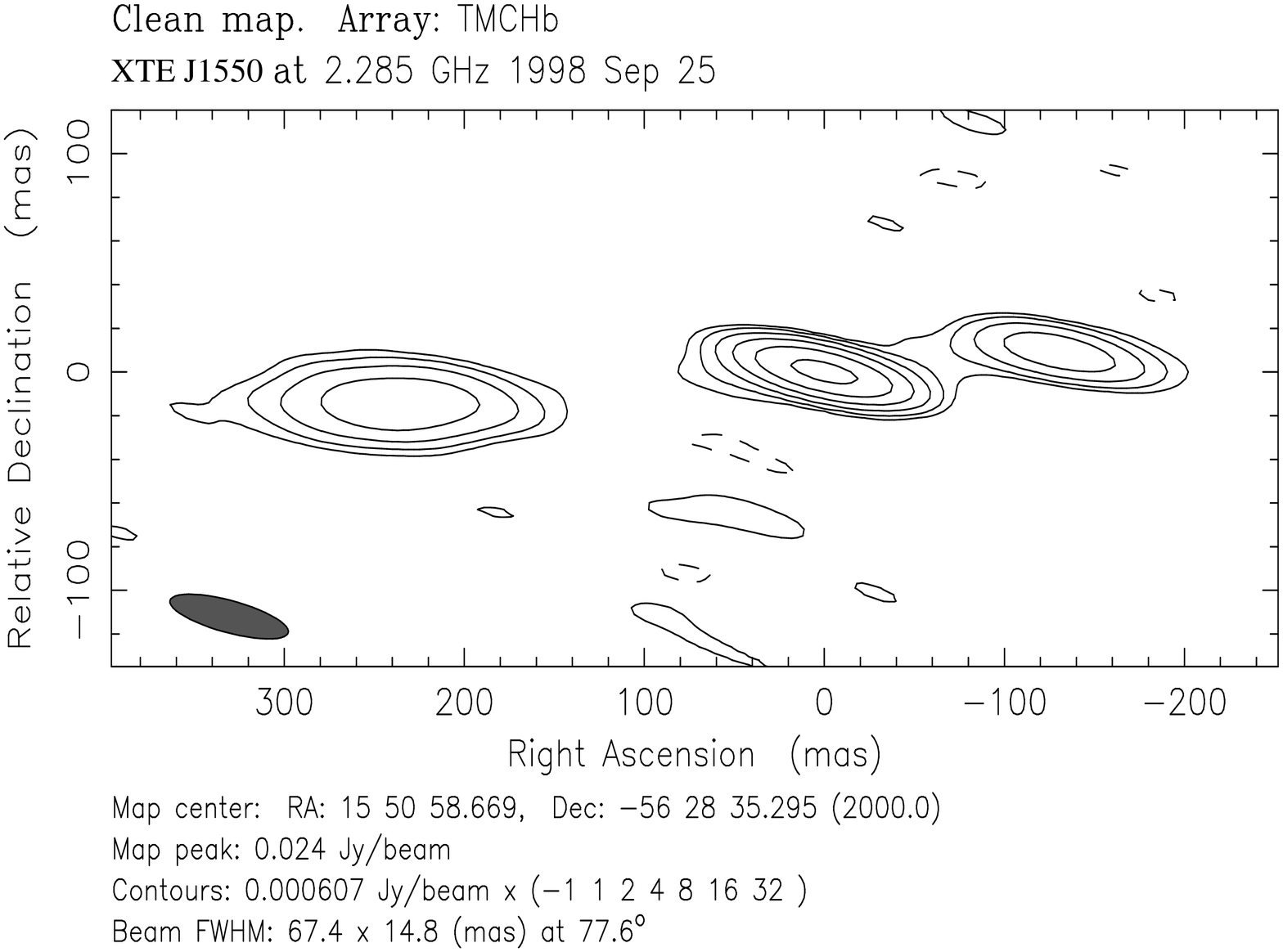}}
\caption{VLBI images at 2.29~GHz from 1998 Sept 24 (MJD~51080.7, top) and
1998 Sept 25 (MJD~51081.7, bottom). The zero coordinate is arbitrary.
The beam is shown in the lower left-hand corner in each image.
}

\end{figure}

Figure~1 shows the resulting VLBI images at 2.29~GHz from
   1998 Sept 24 in the top panel and from 1998 Sept 25 in the bottom panel.
Phase-referenced observations were not performed on either day so the zero
   coordinate in the images is arbitrary.
The flux densities (with un\-certainties of 10--20\%) of the components are
summarized in Table~\ref{tabfd}.

\begin{table}
\caption{Flux densities at 2.29~GHz}
\label{tabfd}
\begin{tabular}{lccccc} \hline
       & \multicolumn{3}{c}{Flux density (mJy)} & & \\
Date        &  East  &  Central   & West & \vline & Total \\
1998 Sept 24 &  71 &    --      &  20   & \vline & 91 \\
1998 Sept 25 &  19    & 25      &   8 & \vline  &  52 \\ \hline
\end{tabular}
\end{table}

The image from 1998 Sept 25 clearly shows an additional component compared to
    the image on 1998 Sept 24.
As the image center is arbitrary, it is not obvious how to relate the components
    seen in the two images.
However, two-point spectral indices derived from Australia Telescope Compact
    Array data at the time of the radio flare show a flattening just prior to
    the first VLBI observation from about $\alpha=-0.45$ to $-0.1$ ($S_{\nu}
    \propto \nu^{\alpha}$) over the interval 4.8--8.6~GHz \cite{hanni1}.
This is very similar to the behavior seen in the 4.8--8.6~GHz two-point spectral
   index of GRO\,J1655$-$40 during the 1994 jet ejection episodes \cite{hanni2}
   and strongly suggests that a new optically thick outburst occurred.
If this interpretation is correct, then we associate the central component
   in the bottom panel with the flaring core and thus
   the two components in the top panel, the ejecta, can be
   associated with the two outermost components in the bottom panel.
This interpretation is reinforced by the 1998 Sept 25 single-baseline 8.4~GHz
data    which shows a new unresolved component not present in the 1998 Sept 24
data. 
This means that in one day the ejecta have moved apart by
approximately 115~mas.
At an estimated distance of 3.5--5~kpc (McIntyre, in preparation)
   this implies a    separation velocity $>2c$ for the ejecta, which makes
   XTE\,J1550$-$564 the    fifth Galactic source to exhibit apparent
   superluminal motion.

\section{Conclusions}

We have presented two VLBI images of XTE\,J1550$-$564 from the time of the
1998 September outburst and showed that the source evolved over a period of
24~hours. The two outermost components seen in the 1998 Sept 25 image are
associated with the two components in the 1998 Sept 24 image, implying an
apparent separation velocity of $>2c$.

\acknowledgements
The Australia Telescope Compact Array and the Mopra Telescope are part of the
Australia Telescope which is funded by the Commonwealth of Australia for
operation as a National Facility managed by CSIRO. We thank the staff
of both the University of Tasmania, Hobart, and the Canberra Deep Space
Communications Complex, Tidbinbilla. DH acknowledges the support of a PPARC
postdoctoral research grant to the University of Southampton.

\end{article}
\end{document}